\documentclass[conference]{IEEEtran}
\IEEEoverridecommandlockouts
\usepackage[hidelinks]{hyperref}
\usepackage{graphicx} 
\usepackage{enumitem}
\usepackage{amsmath,amssymb,amsfonts,eucal}
\usepackage{algorithmic}
\usepackage{graphicx,float,tabularray,supertabular}
\graphicspath{{./figs/}}
\usepackage{textcomp}
\usepackage{xcolor}
\usepackage[square,numbers,sort&compress]{natbib}
\usepackage{ieeemacros}
\usepackage{soul}
\usepackage{optdefi}
\def\BibTeX{{\rm B\kern-.05em{\sc i\kern-.025em b}\kern-.08em
    T\kern-.1667em\lower.7ex\hbox{E}\kern-.125emX}}
\begin{document}

\title{
A Stackelberg Game of Demand Response from the Aggregator's Perspective*
\thanks{* Supported by Petchra Prajom Klao Research Scholarship}
}

\author{
\IEEEauthorblockN{1\textsuperscript{st} Seangleng Khe}
\IEEEauthorblockA{
  \textit{(1): The Joint Graduate School} \\
  \textit{of Energy and Environment,} \\
  \textit{King Mongkut's University}\\
  \textit{of Technology Thonburi, and} \\
  \textit{(2): Center of Excellence on} \\
  \textit{Energy Technology and Environment (CEE),} \\
  \textit{Ministry of Higher Education, Science,} \\
  \textit{Research and Innovation (MHESI),} \\
  \textit{Bangkok, Thailand}
  kheseangleng@gmail.com
}
\and
\IEEEauthorblockN{2\textsuperscript{nd} Parin Chaipunya}
\IEEEauthorblockA{
  \textit{(1): The Joint Graduate School} \\
  \textit{of Energy and Environment, and} \\
  \textit{(2): Department of Mathematics,} \\
  \textit{King Mongkut's University}\\
  \textit{of Technology Thonburi,} \\
  \textit{Bangkok, Thailand} \\
  parin.cha@kmutt.ac.th
}
\and
\IEEEauthorblockN{3\textsuperscript{rd} Athikom Bangviwat}
\IEEEauthorblockA{
  \textit{The Joint Graduate School} \\
  \textit{of Energy and Environment,} \\
  \textit{King Mongkut's University}\\
  \textit{of Technology Thonburi,} \\
  \textit{Bangkok, Thailand} \\
  athikom.bangviwat@outlook.com
}
}
\maketitle

\begin{abstract}
  In this paper, we investigate on the modeling of demand response activities between the single aggregator and multiple participating consumers.
  The model incorporates the bilevel structure that naturally occurs in the information structure and decision sequence, where the aggregator assumes the role of a leader and the participating consumers play the role of followers.
  The proposed model is demonstrated to be effective in load control, helping the aggregator to meet the target reduction while the consumers pay cheaper electricity bill.
\end{abstract}


\section{Introduction}

Demand response refers to the mechanism which drives the end-consumers to change their electricity consumption behaviors \cite{MORALESESPANA2022122544, PATERAKIS2017871, oconnell2014benefits}.
This program has been employed in different sectors including industrial, commercial building, and residential sectors.
\citet{Amir6719589} provided a case study regarding the operation of distribution networks in the residential sector.
This study reported the significance of demand response for network operation, reliability, and responsive heating and ventilation systems.
Similarly, \citet{amin2023demand} identified the roles and benefits of demand response for Australian regional distribution networks.
Usually, a demand response program associated with the Time-of-Use (ToU) tariff-based scheme provided the desired productivity --- 25\% of the loads reduction, improvement of the network voltage profiles, and the grid relaxation.
Different technologies including smart meters are introduced \cite{sameer7931852}, and incentivized programs are applied to achieve a demand response program \cite{EID201615, Yu8972269, digiorgio2014}.
\citet{yu2016} embedded real-time pricing within a demand response program to observe the benefits in the energy system.
The study showed significant advantages enabling load shifting and reducing the energy bill by up to 20.7\% per day for the consumers.
\citet{sanguansuttigul2024bilevel} designed a bilevel optimization model to observe the relation between an energy provider and multiple consumers in the energy system.
The provider acts as a leader aiming to maximize the profit while the consumers act as followers aiming to minimize their electricity bill.
The authors of \cite{sanguansuttigul2024bilevel} had included a storage facility in the system.
To solve the proposed model, the authors utilized the Karush-Kunh-Tucker (KKT) conditions to transform it into a single-level optimization problem, and the SOS-1 approach was applied to deal with complementarity constraints.
The results illustrated that load shifting was achieved. 
The study by \citet{TIWARI2025101671} proposed a bilevel optimization to capture the relation between a distribution system operator (DSO) and an aggregator.
The DSO plays as a leader and the aggregators play as followers.
The leader has multiple objectives including maximizing the operation cost while minimizing network energy loss and peak load at the point of common coupling.
The aggregator acts as followers aiming to maximize incentives to the end-consumers while minimizing their discomfort that occurs in the process. 
The authors then tested the proposed models with the IEEE 25-bus unbalanced distribution system.
The result illustrated that the end-consumers can save 6\% in electricity bill, 20\% improvement in peak-of-average ratio for DSO, and able to reduced loss of 22\%.

\subsection*{Contribution}
In this paper, the study is driven by the fact that a utility company gives a mission to the aggregator to induce the elctricity consumption during on-peak hours.
The aggregator will pass the information to his perspective consumers to shift the load.
We will propose a model to capture this interconnection between the aggregator and the consumers.
The observation shall be what decision should the aggregator makes so that the consumers shall respond in a way he prefers in order to achieve the mission set by the utility company, and he shall obtain the commission most which will be determined based on his performance of electricity reduction.
We will model the observation using a Stackelberg game (also known as bilevel optimization).
The game was first introduced by \citet{stackelberg34} and it gains its popularity due to its special structure.
This technique reflects a hierarchical decision-making process, where one decision influences another decisions, and considers the interactions between stakeholders inside the designated system.
The main constributions of this paper are as follows:
\begin{enumerate}[label=\arabic*., leftmargin=*]
  \item A demand-response one-leader,~$N$-follower optimistic bilevel is proposed to capture the interactions between the aggregator and the consumers in inducing the electricity consumption during on-peak hours.
  \item A dissatifaction function is defined to observe the behaviours of the consumers towards the call from the aggregator.
  \item A numerical simulation has been delivered to testify the proposed model.
  As a result, the model enables load control by active consumers.
\end{enumerate}

\section{Bilevel formulation}
\subsection{Problem description} 

In the electricity market, there are different stakeholders including producers, utility companies, aggregators, and consumers \cite{ALAMOUSH2024114074, Maharjan6464552}.
In this study, the framework is induced to interactions between two stakeholders --- one aggregator and multiple consumers.
The general goal is to shift the high consumption during the on-peak hours to the off-peak hours.
The primary strategy is to set a higher tariff~$\onprice$ to the on-peak hours and a lower tariff~$\offprice$ to the off-peak hours.
This is sometimes insufficient and a more active demand response program has to be introduced.
In such a program, the aggregator will be notified by the utility company to control the electricity consumption of its consumers during the on-peak hours.
The aggregator will pass this piece of information to the consumers.
Since there are many consumers under his control, the aggregator needs to decide which consumers should he call and not to call.
He also decides how much electricity each consumer should he call to shift their load based on the baseline electricity consumption, denoted by~$\baseline_{\consumer}$, of consumer~$\consumer$.
In this load-shedding operation, the aggregator shall receive the commission based on its performance.
Therefore, the aggregator must try its best to cut the electricity consumption during on-peak hours.
For the consumers, when they receive the call from the aggregator they can either fully or partially comply with the call.
By fully complying, consumers will be able to shift the electricity consumption as requested from on-peak hours, as partially complying refers to the situation when the consumers cannot reduce the exact amount of electricity request.
In this sense, the consumers have options to react to the call when possible.
Similarly, the consumers would receive rewards based on their performance.

\subsection{Concept of the proposed model}
In this study, the one-leader,~$N$-follower optimistic bilevel programming is developed to consider the interactions between two stakeholders --- aggregator and consumers.
The aggregator plays as a leader and the consumers play as followers in the proposed model.
The aggregator when receiving the mission from the utility to reduce the electricity consumption during on-peak hours, passes the information to the end-consumers.
The strategy of the leader is defined by the set of electricity calls to the consumers and the amount of electricity reduction sent to each participating consumer.
The benefit of the leader is obtained by commissioning when calling the consumers.
The consumers who participate in the mission would be incentivized and rewarded when they shift the electricity consumption to off-peak hours.
The results of the problem shall be the optimal load control achieved by the consumers regarding the call for electricity reduction as the followers would choose the best strategic electricity call to respond and the aggregator shall select his best strategy providng that each consumer optimally responds to his decision.
\begin{figure}
  \centering
  \includegraphics[scale=0.4]{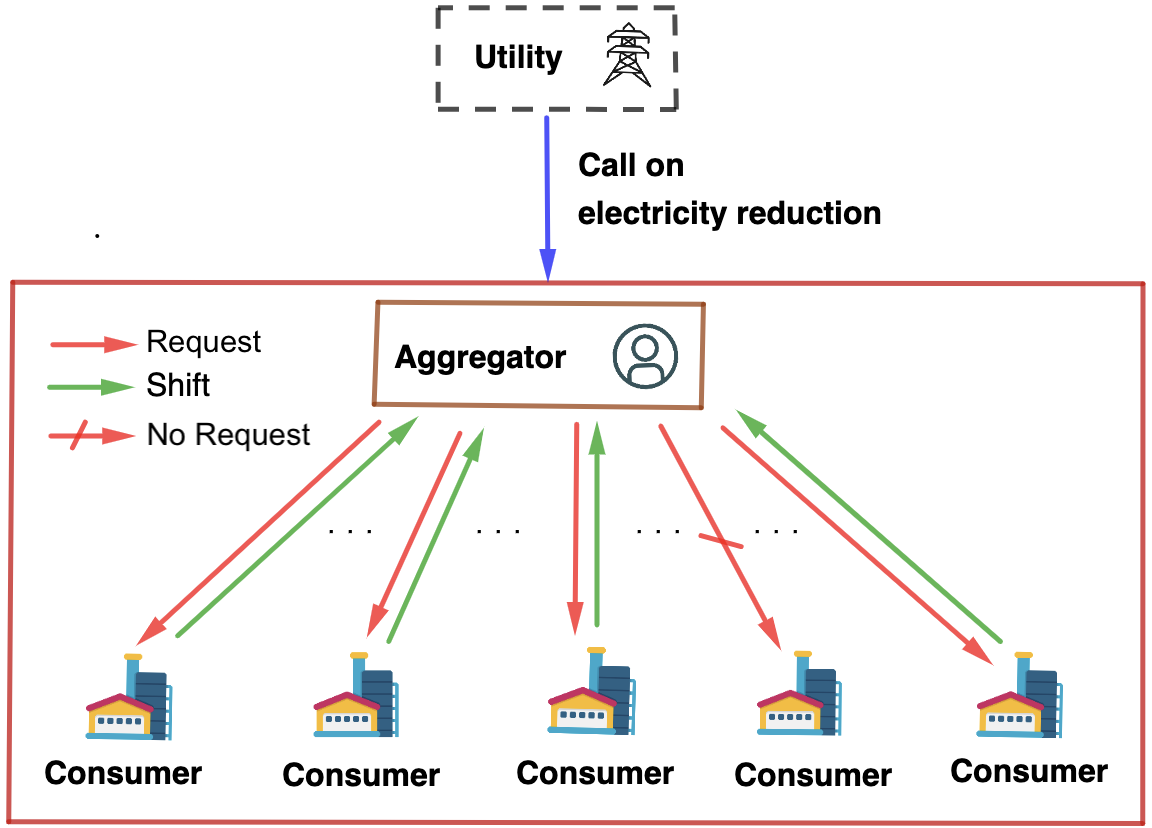}
  \caption{The framework of the study.}
\end{figure}

\subsection{Follower (consumers) model} 
The consumers play as followers in this study.
They can be residential, commercial or industrial.
The total number of consumer is~$\totalConsumers$ and each consumer~$\consumer \in \{1,\dots,\totalConsumers\}$.
There is no guarantee that all consumers will be called upon to join the mission of electricity reduction during the on-peak hours.
On the other hand, the called consumers can decide to take part in the mission or not depending on their availabitity to shift.
The consumers who are willingly to join the mission shall be rewarded which depends on the portion~$\shift_{\consumer} \in [0,1]$ he can make.
The objective function of the consumer~$\consumer$ is defined by
\begin{equation} \label{eq:objFollower}
  \onprice \baseline_{\consumer} (1 - \shift_{\consumer}) + \offprice \baseline_{\consumer} \shift_{\consumer} + \dissat_{\consumer}(\shift_{\consumer}) - \rewardfactor \shift_{\consumer} \text{\baseline}_{\consumer} (\onprice - \offprice),
\end{equation}
where~$\dissat_{\consumer}$ and~$\rewardfactor$ denote the dissatisfaction function of the consumer~$\consumer$ and the reward, respectively.
After a simple calculation, we could simplify the objective function~\eqref{eq:objFollower} as
\begin{equation}
  \onprice \baseline_{\consumer} - (1 + \rewardfactor) \baseline_{\consumer} \shift_{\consumer} ( \onprice - \offprice ) + \dissat_{\consumer}(\shift_{\consumer}).
\end{equation}

In equation~\eqref{eq:objFollower}, the first two terms represent the electricity bill that each consumer needs to pay.
The third term, which describes his dissatifaction, is modeled using a quadratic function
\begin{equation}
  \dissat_{\consumer}(\shift_{\consumer}) = a_{\consumer} \shift_{\consumer}^2 - b_{\consumer} \shift_{\consumer},
\end{equation}
where~$a_{\consumer}, b_{\consumer}$ are positive, consumer-dependent parameters reflecting different characteristics.
The last expression of~\eqref{eq:objFollower} is the reward each consumer~$\consumer$ shall obtain per shifted unit.

In the follower's model, the consumer~$\consumer$ would select their best strategy~$\shift_{\consumer}$ (the percentage of amount to shift electricity) so that they can save the bill and get more rewards.
In summary, the consumer~$\consumer$'s optimization problem reads
\begin{optim}
  \setObj{\min}{\shift_{\consumer}}{ a_{\consumer} \shift_{\consumer}^2 - (1 + \rewardfactor) \baseline_{\consumer} \shift_{\consumer} ( \onprice - \offprice ) - b_{\consumer} \shift_{\consumer} + \onprice \baseline_{\consumer} } \label{eq:FollowerProbFirst}
  \addIneqCons{\shift_{\consumer} \baseline_{\consumer}}{\call_{\consumer}} \label{eq:followerCons}
  \addInclCons{\shift_{\consumer}}{[0,1], \quad \forall i= 1, \ldots, \totalConsumers.}\label{eq:FollowerProbLast}
\end{optim}
This optimization problem is actually parametrized by the call~$\call_{\consumer}$, which is external to the consumer's scope.
Also note that~\eqref{eq:followerCons} ensures that the reduced amount does not exceed the initial amount~$c_{i}$ asked by the aggregator.
This is justified by the fact that the consumer will not get an extra reward beyond the threshold value~$\call_{\consumer}$.

\subsection{Leader (aggregator) model} 
The aggregator plays as a leader in the study.
After recieving the mission from the utility to reduce the electricity consumption~$\target$ during on-peak hours, he shall call on the consumer~$\consumer$ to shift the consumption.
He shall decide which consumer~$\consumer$ should he call and the reduction amount~$\call_{\consumer}$ to ask for.
The objective function of the aggregator is given by
\begin{equation} \label{eq:objLeader}
  \rate (\onprice - \offprice) \sum_{\consumer} \shift_{\consumer} \baseline_{\consumer} - \frac{\fairfactor}{\totalConsumers} \sum_{\consumer} (\midcall - \call_{\consumer})^2,
\end{equation}
constructed as the commision he received weigted by the fairness of the calls.

In Equation~\eqref{eq:objLeader}, the first term represents the commission based on his performance with the commission rate~$\rate$.
While the second term refers to the fairness of the aggregator, which is weigted by the parameter~$\fairfactor$.
Note that the fairnes term is actually measures a squared-deviation of the calls from their average~$\midcall = \frac{1}{N}\sum_{i}\call_{\consumer}$.

In addition to the presented objective function, the aggregator is also bound to the following constraints
\begin{subequations}
  \begin{align*}
    &\sum_{\consumer} \call_{\consumer} = \target, \\
    &\call_{\consumer} \leq \baseline_{\consumer}, \quad (\forall \consumer = 1,\dots,\totalConsumers).
  \end{align*}
\end{subequations}
These constraints guarantee that the total call sums up to the reduction target and each call respects the consumer's baseline consumption.

\subsection{The Stackelberg game} 

Putting together the followers' and leader's optimization problems, we obtain a Stackelberg game with~$\totalConsumers$ followers as follows
\begin{subequations} 
  \begin{empheq}[left=\empheqlbrace \ ]{align}
    \max_{\call_{\consumer}} \quad & \rate (\onprice - \offprice) \sum_{\consumer} \shift_{\consumer} \baseline_{\consumer} - \frac{\fairfactor}{\totalConsumers} \sum_{\consumer} (\midcall - \call_{\consumer})^2 \label{eq:BiProgFirst}\\
    \subjto \quad  & \sum_{\consumer} \call_{\consumer} = \target \\
    & \call_{\consumer} \le \baseline_{\consumer} \\
    & \shift_{\consumer} \in \mathcal{S}(\call_{\consumer}), \label{eq:BiProgLast}
  \end{empheq}
\end{subequations}
where~$\mathcal{S}(\call_{\consumer})$ is the solution set of the follower~$\consumer$'s problem~\eqref{eq:FollowerProbFirst}--\eqref{eq:FollowerProbLast}.

One efficient technique to solve the problem~\eqref{eq:BiProgFirst}--\eqref{eq:BiProgLast} is based on a single-level transformation obtained by replacing the followers' problems with their corresponding Karush-Kuhn-Tucker (KKT) conditions \cite{dempe-dutta-12, dempe2007optimality}.
This single-level problem is in the form of \emph{Mathematical Program with Complementarity Constraints} (MPCC), which is still a challenging class of problem nevertheless.

\section{Results and discussions}

Numerical simulations are performed to validate the proposed model.
We use {\ttfamily Julia programming language} with {\ttfamily JuMP package} and {\ttfamily Gurobi solver} for the simulations.
The data used here are synthetic data that were generated based on literature reviews.

In our simulations, we consider~$\totalConsumers = 10$ consumers participated in the demand response program.
The electricity baseline~$\baseline_{\consumer}$ of each consumer~$\consumer=1,\dots,\totalConsumers$ is set between 80 and 150 kWh except for one consumer that his baseline is extremely high at 800 kWh.
In this framework, the on-peak hour is between 09h00 and 21h00, while the off-peak hour is between 21h00 and 09h00 of the next day.
We consider two scenarios, where to target calls are 800 kWh and 1 500 kWh, respectively.

\subsection{Optimal solution for the aggregator}

\begin{figure}[h]
  \centering
  \includegraphics[width=.47\textwidth]{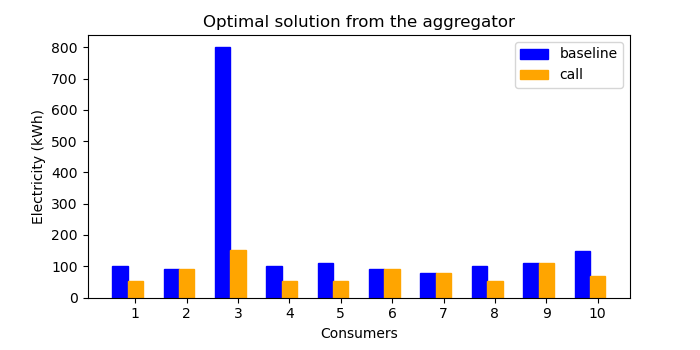}
  \caption{The optimal solution of the aggregator for the target is 800 kWh.}
  \label{fig:R is 800 big a}
\end{figure}

\begin{figure}[h]
  \centering
  \includegraphics[width=.47\textwidth]{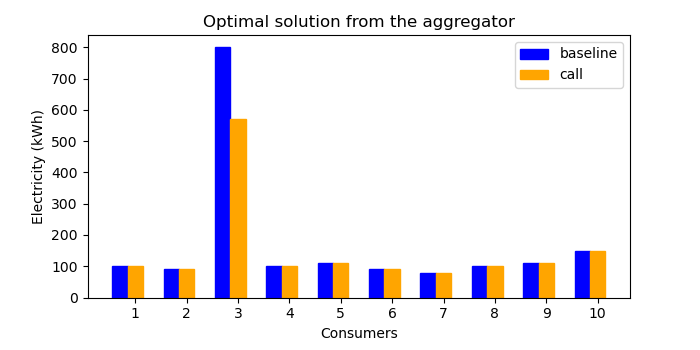}
  \caption{The optimal solution of the aggregator for the target is 1 500 kWh.}
  \label{fig:R is 1500 big a}
\end{figure}

For the case of~$R=$ 800 kWh. 
The aggregator calls on all consumers to shift the load with the same amount of electricity.
Yet when we modify the dissatisfaction function so that it has large coefficient especially~$a_{\consumer}$, we see some differences.
Instead of calling on consumers with the similar amount, the aggregator carefully assigns the amount to each consumer.
For instance, he calls on the consumers with large dissatisfaction function less amount even if they have high electricity consumption, and other consumers with small~$a_{\consumer}$  high amount \figurename~\ref{fig:R is 800 big a}.
For the case of~$R$ = 1 500 kWh, the aggregator calls on the easy-consumers to shift very high amount for example consumer 3 and fewer amount from hard-consumers as shown in \figurename~\ref{fig:R is 1500 big a}.
As a result, The success rate is high in terms of percentage.
With the reduction of 800 kWh, the aggregator achieves approximately 84\%, and for the target of 1 500 kWh, he achieves approximately 73\%.
As defined in~\eqref{eq:objLeader}, the commission is propotional to the amount of shifted load.

\subsection{Optimal solution for the consumers}

\begin{figure}[h]
  \centering
  \includegraphics[width=.47\textwidth]{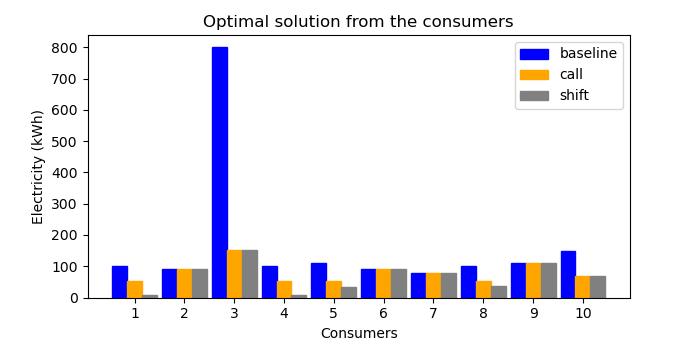}
  \caption{The optimal solution of the consumers for the target is 800 kWh.}
  \label{fig:R is 800 big a Consumer}
\end{figure}

Recieving the request of the aggregator, all consumers optimally comply with the call.
For small coefficients of the dissatisfaction function, the consumers fully comply with call and switch their operation to off-peak hours.
Similar to the previous argument, when the coefficients especially~$a_{\consumer}$ are large, the response is distinctive.
In \figurename~\ref{fig:R is 800 big a Consumer}, some consumers fully comply with call and shift their operation to the off-peak hours like consumer 2, 6, 7, and 9.
Consumer 2 which has the highest consumption is called to shift 152.4 kWh (less than his consumption) and he fully complies with the call and shift the exact requested amount.
For the other consumers, e.g. consumer 1 with the baseline of 90 kWh is called to shift 52.4 kWh yet he manages to shift only 7.55 kWh.
This means that he partially complies with the call.
Similar argument is obtained, when the $R=$ 1 500 kWh.
The consumers who fully comly with the call, still fully comply and the consumers who partially comply, still partially comply with the call.
Besides, the consumers who are willingly complying with the call shall obtain more reward and pay fewer electricity bill.
For instance, consumer 2 recieves most reward and pays electricity at off-peak rate.
Consumer 3 also recieves the most reward by fully complying with the call and he pay less for electricity bill.

\begin{figure}[h]
  \centering
  \includegraphics[width=.47\textwidth]{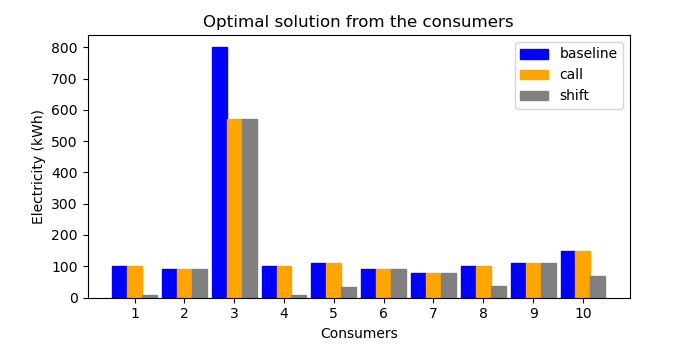}
  \caption{The optimal solution of the consumers for the target is 1 500 kWh.}
  \label{fig:R is 1500 big a Consumer}
\end{figure}

\subsection{Fairness and the commission of the aggregator }

\begin{figure}[h]
  \centering
  \includegraphics[width=.47\textwidth]{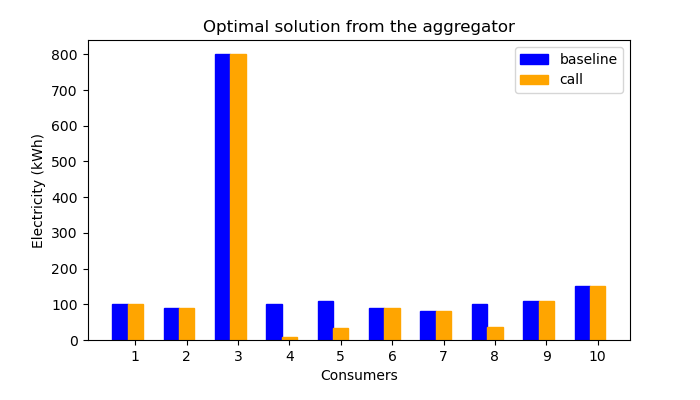}
  \caption{The optimal solution of the consumers when the target is 1,500 kWh and the fair term is removed.}
  \label{fig:R is 1500 big a Consumer no fair}
\end{figure}

As discussed earlier that the commission is propotional to the amount of elctricity reduction.
Yet we take more investigation the relation between the fairness and the commission.
Let us recall that we incorporate the fairness term into the leader's objective function to force it to spread the call as equally as possible under the performance history of each consumers.
Without the fairness term, this fairness aspect is not taken into consideration.
To emphasize this unfair behavior, a simulation for the reduction target of 1 500 kWh is considered again without the fairness term (\emph{i.e.}~$\fairfactor = 0$).
The optimal solution of the aggregator is shown in the \figurename~\ref{fig:R is 1500 big a Consumer no fair}, where one could see that the call is strongly focused on the consumer 3.
It is notable that although the mission of aggregating load reduction is achieved at~$88 \%$ the value variance~$\frac{1}{N}\sum_{\consumer}(\midcall - \call_{\consumer})^{2}$ is much higher than the results with fairness term presented earlier.

With more simulations performed, we observe, following our expectation, that the variance (representing the unfairness) increases as~$\fairfactor$ decreases.

\section{Conclusions}

An innovative bilevel programming has been modeled to deal with the energy consumption reduction during on-peak hours.
The simulation result demonstrates the validity of the model which is able to induce the electricity consumption during peak hours and achieve load shift in energy management.
The aggregator makes the call regarding the actual consumption baseline and the availability based on the dissatisfaction of the consumers.
To be able to make an effective call, the aggregator observes every consumer whether he can fully comply and is satisfied with it.
From the consumers' side, their participations depend largely on their willingness.
If the consumers think the call is larger than what they could bear, they partially comply with the recieved calls, or in some cases, completely ignore the call.


\begin{thebibliography}{17}
\providecommand{\natexlab}[1]{#1}
\providecommand{\url}[1]{#1}
\csname url@samestyle\endcsname
\providecommand{\newblock}{\relax}
\providecommand{\bibinfo}[2]{#2}
\providecommand{\BIBentrySTDinterwordspacing}{\spaceskip=0pt\relax}
\providecommand{\BIBentryALTinterwordstretchfactor}{4}
\providecommand{\BIBentryALTinterwordspacing}{\spaceskip=\fontdimen2\font plus
\BIBentryALTinterwordstretchfactor\fontdimen3\font minus \fontdimen4\font\relax}
\providecommand{\BIBforeignlanguage}[2]{{%
\expandafter\ifx\csname l@#1\endcsname\relax
\typeout{** WARNING: IEEEtranN.bst: No hyphenation pattern has been}%
\typeout{** loaded for the language `#1'. Using the pattern for}%
\typeout{** the default language instead.}%
\else
\language=\csname l@#1\endcsname
\fi
#2}}
\providecommand{\BIBdecl}{\relax}
\BIBdecl

\bibitem[Morales-España et~al.(2022)Morales-España, Martínez-Gordón, and Sijm]{MORALESESPANA2022122544}
\BIBentryALTinterwordspacing
G.~Morales-España, R.~Martínez-Gordón, and J.~Sijm, ``Classifying and modelling demand response in power systems,'' \emph{Energy}, vol. 242, p. 122544, 2022. [Online]. Available: \url{https://www.sciencedirect.com/science/article/pii/S0360544221027936}
\BIBentrySTDinterwordspacing

\bibitem[Paterakis et~al.(2017)Paterakis, Erdinç, and Catalão]{PATERAKIS2017871}
\BIBentryALTinterwordspacing
N.~G. Paterakis, O.~Erdinç, and J.~P. Catalão, ``An overview of demand response: Key-elements and international experience,'' \emph{Renewable and Sustainable Energy Reviews}, vol.~69, pp. 871--891, 2017. [Online]. Available: \url{https://www.sciencedirect.com/science/article/pii/S1364032116308966}
\BIBentrySTDinterwordspacing

\bibitem[O'Connell et~al.(2014)O'Connell, Pinson, Madsen, and O'Malley]{oconnell2014benefits}
N.~O'Connell, P.~Pinson, H.~Madsen, and M.~O'Malley, ``Benefits and challenges of electrical demand response: A critical review,'' \emph{Renewable and Sustainable Energy Reviews}, vol.~39, pp. 686--699, 2014.

\bibitem[Safdarian et~al.(2016)Safdarian, Fotuhi-Firuzabad, and Lehtonen]{Amir6719589}
A.~Safdarian, M.~Fotuhi-Firuzabad, and M.~Lehtonen, ``Benefits of demand response on operation of distribution networks: A case study,'' \emph{IEEE Systems Journal}, vol.~10, no.~1, pp. 189--197, 2016.

\bibitem[Amin et~al.(2023)Amin, Shah, Amjady, Hasan, Tayab, and Islam]{amin2023demand}
B.~M. Amin, R.~Shah, N.~Amjady, K.~Hasan, U.~B. Tayab, and S.~Islam, ``Demand response on the operation of regional distribution network: An australian case study,'' in \emph{2023 IEEE Power \& Energy Society General Meeting (PESGM)}, 2023.

\bibitem[Hoosain and Paul(2017)]{sameer7931852}
M.~S. Hoosain and B.~S. Paul, ``Smart homes: A domestic demand response and demand side energy management system for future smart grids,'' in \emph{2017 International Conference on the Domestic Use of Energy (DUE)}, 2017, pp. 285--291.

\bibitem[Eid et~al.(2016)Eid, Koliou, Valles, Reneses, and Hakvoort]{EID201615}
\BIBentryALTinterwordspacing
C.~Eid, E.~Koliou, M.~Valles, J.~Reneses, and R.~Hakvoort, ``Time-based pricing and electricity demand response: Existing barriers and next steps,'' \emph{Utilities Policy}, vol.~40, pp. 15--25, 2016. [Online]. Available: \url{https://www.sciencedirect.com/science/article/pii/S0957178716300947}
\BIBentrySTDinterwordspacing

\bibitem[Yu et~al.(2019)Yu, Hong, Zhang, Jiang, Huang, Wei, Wang, and Liang]{Yu8972269}
M.~Yu, S.~H. Hong, X.~Zhang, J.~Jiang, X.~Huang, M.~Wei, K.~Wang, and W.~Liang, ``Incentivizing strategy for demand response aggregator considering market entry criterion: A game theoretical approach,'' in \emph{2019 IEEE 17th International Conference on Industrial Informatics (INDIN)}, vol.~1, 2019, pp. 1077--1082.

\bibitem[Di~Giorgio and Liberati(2014)]{digiorgio2014}
\BIBentryALTinterwordspacing
A.~Di~Giorgio and F.~Liberati, ``Near real-time load shifting control for residential electricity prosumers under designed and market-indexed pricing models,'' \emph{Applied Energy}, vol. 128, pp. 119--132, 2014. [Online]. Available: \url{http://dx.doi.org/10.1016/j.apenergy.2014.04.032}
\BIBentrySTDinterwordspacing

\bibitem[Yu and Hong(2016)]{yu2016}
M.~Yu and S.~H. Hong, ``A real-time demand-response algorithm for smart grids: A stackelberg game approach,'' \emph{IEEE Transactions on Smart Grid}, vol.~7, no.~2, March 2016.

\bibitem[Sanguansuttigul et~al.(2024)Sanguansuttigul, Chayawatto, and Chaipunya]{sanguansuttigul2024bilevel}
\BIBentryALTinterwordspacing
P.~Sanguansuttigul, N.~Chayawatto, and P.~Chaipunya, ``A bilevel qp-plp approach to demand response modulation between consumers and a single electricity seller,'' \emph{Science \& Technology Asia}, vol.~29, no.~2, pp. 32--44, 2024. [Online]. Available: \url{https://ph02.tci-thaijo.org/index.php/SciTechAsia/article/view/254626}
\BIBentrySTDinterwordspacing

\bibitem[Tiwari et~al.(2025)Tiwari, Jha, and Pindoriya]{TIWARI2025101671}
\BIBentryALTinterwordspacing
A.~Tiwari, B.~K. Jha, and N.~M. Pindoriya, ``Incentive-based demand response program with phase unbalance mitigation: A bilevel approach,'' \emph{Sustainable Energy, Grids and Networks}, vol.~42, p. 101671, 2025. [Online]. Available: \url{https://www.sciencedirect.com/science/article/pii/S2352467725000530}
\BIBentrySTDinterwordspacing

\bibitem[von Stackelberg(1934)]{stackelberg34}
H.~von Stackelberg, \emph{\BIBforeignlanguage{German}{Marktform und {Gleichgewicht}}}.\hskip 1em plus 0.5em minus 0.4em\relax Berlin: Julius Springer, 1934.

\bibitem[Alamoush et~al.(2024)Alamoush, Ballini, and Ölçer]{ALAMOUSH2024114074}
\BIBentryALTinterwordspacing
A.~S. Alamoush, F.~Ballini, and A.~I. Ölçer, ``Management of stakeholders engaged in port energy transition,'' \emph{Energy Policy}, vol. 188, p. 114074, 2024. [Online]. Available: \url{https://www.sciencedirect.com/science/article/pii/S0301421524000946}
\BIBentrySTDinterwordspacing

\bibitem[Maharjan et~al.(2013)Maharjan, Zhu, Zhang, Gjessing, and Basar]{Maharjan6464552}
S.~Maharjan, Q.~Zhu, Y.~Zhang, S.~Gjessing, and T.~Basar, ``Dependable demand response management in the smart grid: A stackelberg game approach,'' \emph{IEEE Transactions on Smart Grid}, vol.~4, no.~1, pp. 120--132, 2013.

\bibitem[Dempe and Dutta(2012)]{dempe-dutta-12}
S.~Dempe and J.~Dutta, ``\BIBforeignlanguage{English}{Is bilevel programming a special case of a mathematical program with complementarity constraints?}'' \emph{\BIBforeignlanguage{English}{Math. Program.}}, vol. 131, no. 1-2 (A), pp. 37--48, 2012.

\bibitem[Dempe et~al.(2007)Dempe, Dutta, and Mordukhovich]{dempe2007optimality}
S.~Dempe, J.~Dutta, and B.~S. Mordukhovich, ``New necessary optimality conditions in optimistic bilevel programming,'' \emph{Optimization}, vol.~56, no. 5-6, pp. 577--604, 2007.

\end{thebibliography}


\end{document}